\documentclass{iopart}
\usepackage{iopams}
\usepackage{subfigure}
\usepackage{graphicx}
\usepackage{url}

\begin{document}

\title[Determining membrane permeability using video
microscopy]{Determining membrane permeability of giant phospholipid
  vesicles from a series of videomicroscopy images}

\author{Primo\v{z} Peterlin${}^1$, Ga\v{s}per Jakli\v{c}${}^{2,3}$, and
Toma\v{z} Pisanski${}^{2,3}$}

\address{${}^1$University of Ljubljana, Faculty of Medicine, Institute
  of Biophysics, Lipi\v{c}eva~2, Ljubljana, Slovenia,
${}^2$University
  of Ljubljana, Faculty of Mathematics and Physics, Jadranska~19, Ljubljana,
  Slovenia,
${}^3$University of Primorska, Primorska Institute for
  Natural Science and Technology, Muzejski trg~2, Koper, Slovenia}

\ead{primoz.peterlin@mf.uni-lj.si}

\date{\today}


\begin{abstract}  
  A technique for determining the permeability of a phospholipid
  membrane on a single giant unilamellar vesicle (GUV) is described,
  which complements the existing methods utilizing either a planar
  black lipid membrane or sub-micrometre-sized liposomes.  A single
  GUV is transferred using a micropipette from a solution of a
  nonpermeable solute into an iso-osmolar solution of a solute with a
  higher membrane permeability.  Osmotical swelling of the vesicle is
  monitored with a CCD camera mounted on a phase contrast microscope,
  and a sequence of images is obtained.  On each image, the points on
  the vesicle contour are determined using Sobel filtering with
  adaptive binarization threshold, and from these, the vesicle radius
  is calculated with a great accuracy.  From the time-dependence of
  the vesicle radius, the membrane permeability is obtained.  Using a
  test set of data, the method provided a consistent estimate of the
  POPC membrane permeability for glycerol, $P = 1.7\times
  10^{-8}\,\textrm{m/s}$, with individual samples ranging from
  $1.61\times 10^{-8}\,\textrm{m/s}$ to $1.98\times
  10^{-8}\,\textrm{m/s}$.  This value is $\approx 40\%$ lower than the
  one obtained on similar systems.  Possible causes for this
  discrepancy are discussed.
\end{abstract}



\submitto{\MST}

\section{Introduction}
\label{sec:intro}


All living cells are enclosed by a lipid bilayer, which serves as a
barrier for macromolecules and ions.  Small uncharged molecules can,
however, permeate across the membrane without requiring assistance of
transmembrane proteins.  Permeability of a lipid bilayer has
traditionally been measured either electrically on a planar black
lipid membrane \cite{Walter:1986}, or optically on
sub-micrometre-sized liposomes (\emph{e.g.,} large unilamellar
vesicles, LUVs), through osmotic swelling or shrinking of vesicles and
a subsequent change in the light scattering
\cite{DeGier:1993,Paula:1996}.  While these techniques have proven to
be successful, they have two drawbacks.  Firstly, they offer only
indirect insight into the mechanism involved, and secondly, an open
question remains whether the results obtained on the membranes of
sub-micrometre-sized LUVs can be applied onto the membranes of
micrometre-sized giant unilamellar vesicles (GUVs), \emph{i.e.,}
whether the membrane permeability is influenced by its curvature.

In this paper, we propose a complementary technique for measuring the
permeability of a lipid bilayer on GUVs using an analysis of a
sequence of videomicroscopy images.  Processing and analysis of
digital video images acquired by an optical microscope has become an
indispensable tool in biophysics, in particular for the analysis of
the shape of contours of phospholipid vesicles, either free
\cite{Engelhardt:1985,Seul:1991} or aspirated in
a micropipette \cite{Heinrich:2005}.

\section{Description}
\label{sec:description}

\subsection{Experimental setup for image acquisition}
\label{sec:image-acquisition}

In the experiment, a single spherical GUV made from
1-palmitoyl-2-oleoyl-\emph{sn}-glycero-3-phosphocholine (POPC) was
selected, fully aspirated into a glass micropipette with a diameter
exceeding the diameter of the vesicle, and transferred from a solution
of a solute with a very low membrane permeability (\emph{e.g.,}
glucose or sucrose) into an iso-osmolar solution of a solute with a
higher membrane permeability (\emph{e.g.,} glycerol), where the
content of the micropipette was released, and the micropipette was
removed.  Vesicle behaviour is recorded with a CCD camera mounted on
the microscope.  Upon the transfer, the vesicle swells osmotically
until it bursts, whereupon another cycle of swelling is commenced.
The rate of swelling depends on membrane permeability and the
difference of the concentration of the permeable solution outside and
inside the vesicle.  Since the whole sequence of the vesicle response
is recorded, and the vesicle radius can be determined from micrograph
images with a great precision, an estimate for the concentration
difference can be obtained from the vesicle geometry, thus making an
estimate of the membrane permeability possible.  Vesicle transfer from
a solution of a non-permeable into a solution of a permeable solute
rather than vice versa, thus resulting in osmotic swelling rather than
osmotic shrinking, has been chosen since it completely determines the
geometry of a vesicle, \emph{i.e.,} the vesicle is spherical
throughout the experiment, thus making a high-precision measurement of
its radius possible.  The micromanipulation experiment is explained in
detail in \cite{Peterlin:2008}.

The experimental setup consisted of a phase contrast inverted optical
microscope (Nikon Diaphot 200, objective 20/0.40 Ph2 DL; Tokyo, Japan)
with the micromanipulating equipment (Narishige MMN-1/MMO-202; Tokyo,
Japan) and a cooled CCD camera (Hamamatsu ORCA-ER; C4742-95-12ERG;
Hamamatsu, Japan), connected via IEEE-1394 to a PC running Hamamatsu
Wasabi software.  The camera provides $1344\times 1024$ 12-bit
grayscale images.  In the streaming mode, the camera throughput is 8.9
images/s.  Even though custom solutions have been shown to provide a
higher throughput in similar cases \cite{Fujiwara:2007}, the software
supplied by the manufacturer was proved to be adequate for the camera.

\subsection{The model for vesicle inflation and burst}
\label{sec:model}

The basic ideas of the model explaining the swelling-burst cycle will
be presented first.  After the transfer, a spherical vesicle filled
with an impermeable solute is immersed in a solution of a permeable
solute.  Since the concentration of the permeable solute outside the
vesicle ($c_\mathrm{p0}$) exceeds its concentration inside the vesicle
($c_\mathrm{p}$), it tends to diffuse into the vesicle.  During this
phase, the amount of the impermeable solute inside the vesicle
($N_\mathrm{i}$) remains constant, while the amount of the permeable
solute ($N_\mathrm{p}$) increases.  Its influx is accompanied by an
influx of water required to maintain the osmotic balance.  Based on
the known data \cite{Walter:1986}, it is assumed in this model that
the permeability of the membrane for water exceeds all other
permeabilities by several orders of magnitude, and consequently the
osmotic balance is achieved almost instantly.  Consequently, in order
to maintain the osmotic balance, the vesicle volume increases as well.
In an instantaneous event of a vesicle burst, a fraction of the
solution is ejected from the vesicle interior.  During the burst, the
concentration of both solutes ($c_\mathrm{p}$, $c_\mathrm{i}$) inside
the vesicle stays constant, while their amount decreases, the decrease
being proportional to the decrease of the vesicle volume.  The cycle
then repeats itself with new values of the concentrations
$c_\mathrm{p}$ and $c_\mathrm{i}$.  The theory of the osmotic
inflation--burst cycle is described in more detail in
\cite{Koslov:1984,Mally:2002,Peterlin:2008}.

The flux of the permeable solute into the vesicle ($j$) is
proportional to the difference of the concentration of permeable
solute outside ($c_\mathrm{p0}$) and inside ($c_\mathrm{p}$) the
vesicle,
\begin{equation}
  j = P (c_\mathrm{p0} - c_\mathrm{p}) \; ,
  \label{eq:flux-glycerol}
\end{equation}
where $P$ is the permeability of the membrane for the permeable
solute.  The number of its molecules inside the vesicle
($N_\mathrm{p}$) increases with time at a rate:
\begin{equation}
  \frac{dN_\mathrm{p}}{dt} = PA \left( c_\mathrm{p0} -
    \frac{N_\mathrm{p}}{V}\right) \; .
  \label{eq:glycero-rate}
\end{equation}
Both the vesicle volume $V$ and the membrane area $A$ vary with time,
which means that in general, (\ref{eq:glycero-rate}) can only be
solved numerically.

\begin{figure}
  \centering\includegraphics[scale=0.6]{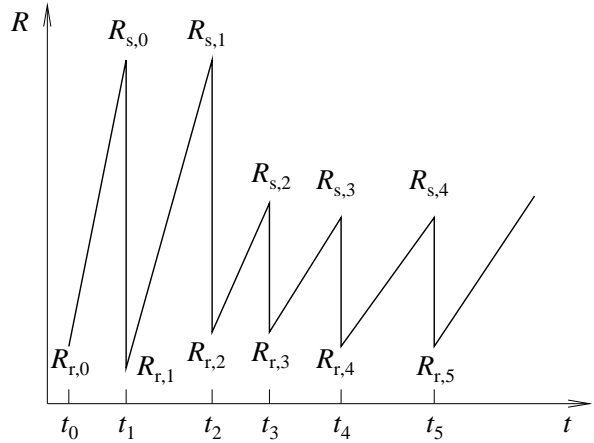}
  \caption{A schematic indication of the vesicle radius $R$ as a
    function of time $t$ upon transfer from a solution of an
    impermeable solute into an iso-osmolar solution of a permeable
    solute.}
  \label{fig:sawtooth}
\end{figure}

For realistic parameter values, however, numerical integration is not
needed.  It is known \cite{Bloom:1991} that the lipid membrane can
only expand by approximately 4\% before its tensile strength is
reached (``stretched'' radius $R_\mathrm{s}$), at which point the
vesicle bursts and ejects its excess volume, and its radius $R$
returns to its ``relaxed'' value $R_\mathrm{r}$
(figure~\ref{fig:sawtooth}).

Such small changes of $R$ can be linearized, leading to a linear
dependence of $V$ in the time between bursts.  A rearrangement of
Eq.~11 in \cite{Peterlin:2008} yields a relation between the rate of
the volume increase in the $n$-th cycle, $\Delta V_n/\Delta t_n$, and
the concentration of the impermeable solute in the vesicle during this
cycle, $N_\mathrm{i}^{(n)}$:
\begin{equation}
  \frac{\Delta V_n}{\Delta t_n} :=
  \frac{V_{\mathrm{s},n} - V_{\mathrm{r},n}}{t_{n+1}-t_n} = 
  \frac{3 P}{R_\mathrm{r,0}} \frac{N_\mathrm{i}^{(n)}}{c_\mathrm{p0}} \; .
  \label{eq:rate}
\end{equation}
We have denoted $V_{\mathrm{r},n} = 4\pi R_{\mathrm{r},n}^3/3$ and
$V_{\mathrm{s},n} = 4\pi R_{\mathrm{s},n}^3/3$.

While $t_n$, $R_{\mathrm{r},n}$ and $R_{\mathrm{s},n}$ can be
determined directly, the value of $N_\mathrm{i}^{(n)}$ at the $n$-th
burst can be calculated.  Since the amount of the impermeable solute
inside the vesicle remains constant during the time between successive
bursts, one can write:
\begin{eqnarray}
  N_\mathrm{i}^{(0)} & = & c_\mathrm{i}^{(0)} V_{\mathrm{r},0}=
  c_\mathrm{i}^{(1)} V_{\mathrm{s},0} \; , \nonumber \\
  N_\mathrm{i}^{(1)} & = & c_\mathrm{i}^{(1)} V_{\mathrm{r},1}=
  c_\mathrm{i}^{(2)} V_{\mathrm{s},1} \; ,  \\
  &\ldots & \nonumber
\end{eqnarray}
From here, a general expression for $c_\mathrm{i}^{(n)}$ can be obtained:
\begin{equation}
  c_\mathrm{i}^{(n)} = c_\mathrm{i}^{(0)} \prod_{j=0}^{n-1}
  \frac{V_{\mathrm{r},j}}{V_{\mathrm{s},j}} \; .
  \label{eq:csn}
\end{equation}
The initial amount of the impermeable solute inside the vesicle is
known, $N_\mathrm{i}^{(0)} = V_{\mathrm{r},0} c_\mathrm{i}^{(0)}$.  We
also know that the solutions inside and outside the vesicle are
iso-osmolar ($c_\mathrm{i}^{(0)} = c_\mathrm{p0}$).  Taking into
account that the relaxed radii of the vesicle are approximately equal
($V_{\mathrm{r},n} \approx V_{\mathrm{r},0}$), and substituting this
into (\ref{eq:rate}), one obtains:
\begin{equation}
  \frac{\Delta V_n}{\Delta t_n} =
  P\, 4\pi R_\mathrm{r,0}^2 \prod_{j=0}^{n-1}
  \frac{V_{\mathrm{r},j}}{V_{\mathrm{s},j}} \; .
  \label{eq:linear}
\end{equation}
Thus, plotting $\Delta V_n/\Delta t_n$ against $4\pi R_\mathrm{r,0}^2
\prod_{j=0}^{n-1} V_{\mathrm{r},j}/V_{\mathrm{s},j}$, one should
obtain a straight line with the slope equal to $P$.

\subsection{Determining vesicle radius from a videomicrograph}

Phase contrast micrographs exhibit a distinct ``halo'', if the index
of refraction of the medium inside does not match the refractive
index of the medium outside \cite{Pluta:AdvLightMicrosc2:halo}.

\begin{figure}
  \centering\includegraphics[scale=0.86]{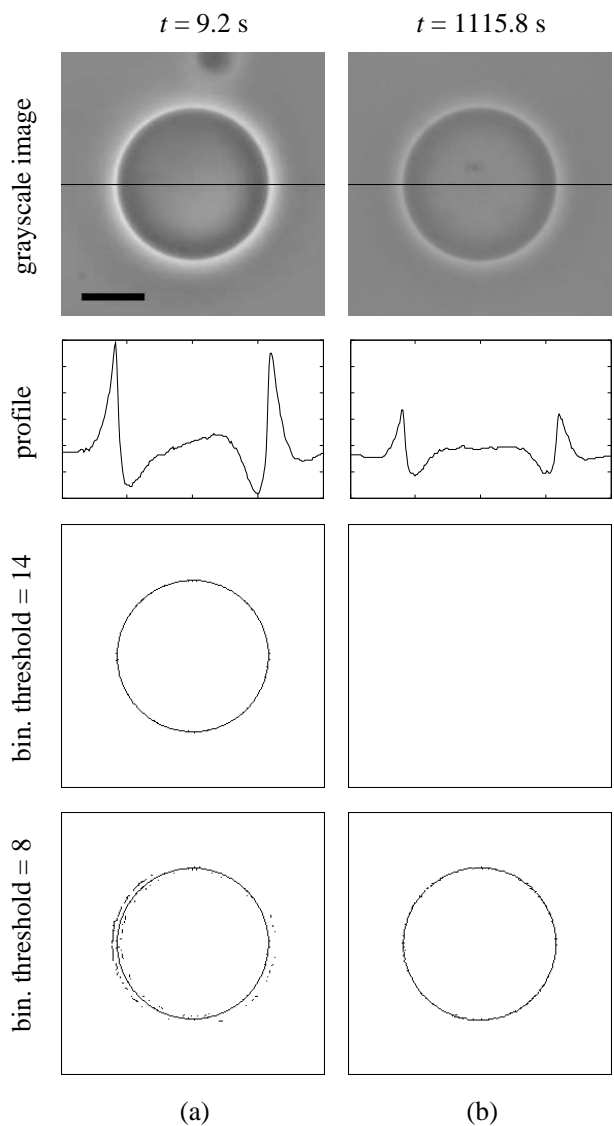}
  \caption{Adaptive Sobel threshold.  As the composition of the
    solution in the vesicle interior approaches the composition of the
    vesicle exterior, the contrast fades from videomicrograph frames
    taken at the beginning of the experiment (a, top row) to frames
    taken towards the end of the experiment (b, top row); the second
    row shows the intensity profile of the vesicle cross-section shown
    in the same scale.  It is thus necessary to dynamically adjust the
    binarization threshold of the Sobel edge operator.  High values of
    threshold, which produce an acceptable image at the beginning of
    the experiment (a, third row), fail to discriminate the edge
    towards the end of the experiment (b, third row).  On the other
    hand, low values of threshold, which are appropriate for the image
    frames towards the end of the experiment (b, bottom row) produce
    stray points at the beginning of the experiment. The algorithm for
    the dynamical adjustment of the binarization threshold selects an
    image corresponding to the one in the third row for (a), and an
    image corresponding to the one in the fourth row for (b). The bar
    in the top left frame represents 20~$\mu$m.}
  \label{fig:adaptive-edge}
\end{figure}

\begin{figure}
  \centering\includegraphics[scale=0.7]{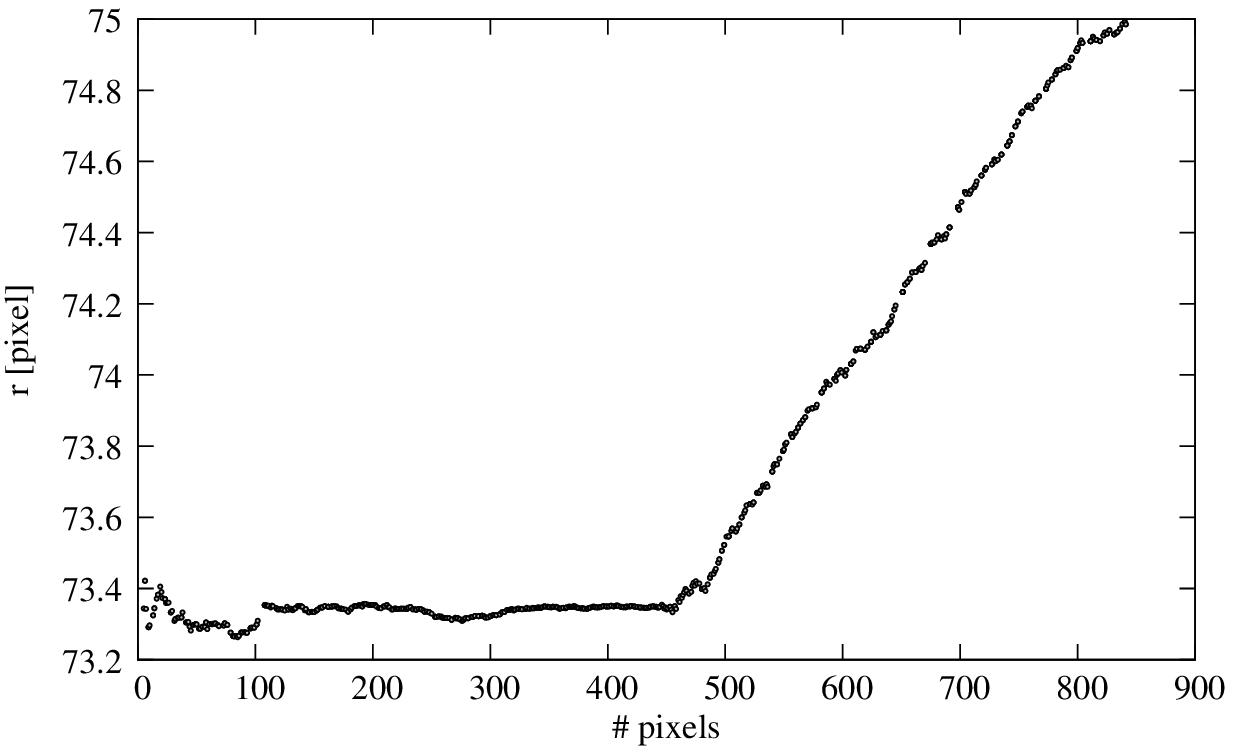}
  \caption{For a given image, the calculated radius of a vesicle
    depends on the number of points taken into consideration.  When
    the binarization threshold is very high, very few points are
    selected and the calculated radius is significantly affected by
    any new point taken into consideration.  In an intermediate
    region, the calculated radius does not depend on the number of
    points.  Above a certain threshold value, spurious points taken
    into consideration affect the calculated radius.  The data plotted
    are for the image plotted in figure~\ref{fig:adaptive-edge}a; the
    interval $[2 \pi R/3, 2 \pi R]$ approximately translates to
    [150,460], corresponding to the interval [9.90, 22.60] for the
    binarization threshold.}
  \label{fig:radius-numpts}
\end{figure}

A Sobel edge detection operator \cite{GonzalezWoods:DIP2e:Ch10} was
used to determine the radius of the vesicle. The choice of threshold
of the binarization applied to the image after the convolution affects
the result in an important way.  Since the contrast of the halo
decreases throughout the time of recording, the binarization threshold
needs to be adjusted dynamically in order to yield the desired result
(figure~\ref{fig:adaptive-edge}).  The binarization threshold was
selected on the basis of the number of points it yields, in a way that
the number of points making up the contour was between $2\pi R/3$ and
$2\pi R$, where $R$ is the running average of previously determined
values, expressed in pixels.  In this interval, the calculated radius
is virtually independent of the number of points taken into
calculation (figure~\ref{fig:radius-numpts}).

A circle with a centre $(x_\mathrm{c},y_\mathrm{c})$ and a radius $R$
is fitted to a set of $n$ points $(x_i,y_i)$ obtained in the previous
step.  A computationally efficient algorithm for fitting a general
conic to a set of data points \cite{Bookstein:1979} was adapted
specifically for fitting ellipses \cite{Fitzgibbon:1999}.  It turns
out that the system obtained in \cite{Fitzgibbon:1999} is unstable,
\emph{e.g.,} the approach does not work for the data, sampled from a
circle.  A more stable and efficient modification has been proposed by
Hal\'{\i}\v{r} and Flusser \cite{Halir:1998}.

Further simplification can be employed for fitting circles, which
reduces the problem to solving a linear $3 \times 3$ system rather
than solving a generalized eigenvalue problem, as required in
\cite{Fitzgibbon:1999}.  If the matrix $\mathbf{A}$ and the vector
$\mathbf{b}$ are constructed as follows:
\begin{eqnarray}
  \mathbf{A} &=& \left[ 
    \begin{array}{lll}
      \sum_i x_i^2   & \sum_i x_i y_i & \sum_i x_i \\
      \sum_i x_i y_i & \sum_i y_i^2   & \sum_i y_i \\
      \sum_i x_i     & \sum_i y_i     & n
    \end{array} \right] \; , \label{eq:matrixa} \\
  \mathbf{b} &=& \left[
    \begin{array}{l}
     \sum_i x_i (x_i^2 + y_i^2) \\ 
     \sum_i y_i (x_i^2 + y_i^2) \\ 
     \sum_i (x_i^2 + y_i^2)
    \end{array}
  \right] \; , \label{eq:vectorb}
\end{eqnarray}
then the required parameters are stored in a vector $\mathbf{v}=(v_1,v_2,v_3)$,
obtained as the solution of a set of linear equations,
\begin{equation}
  \label{eq:vectorv}
  \mathbf{A}\cdot\mathbf{v} = \mathbf{b} \; .
\end{equation}
In particular,
\begin{eqnarray}
  x_\mathrm{c} &=& \frac{1}{2}v_1 \; , \label{eq:xcenter} \\
  y_\mathrm{c} &=& \frac{1}{2}v_2 \; , \label{eq:ycenter} \\
  R &=& \sqrt{x_\mathrm{c}^2 + y_\mathrm{c}^2 + v_3} \; . \label{eq:radius}
\end{eqnarray}

The matrix $\mathbf{A}$ in (\ref{eq:matrixa}) is positive definite,
provided that at least 3 data points are non-collinear.  This can be
easily seen from the factorization $\mathbf{A}= \mathbf{C}^T
\mathbf{C}$, where $\mathbf{C}:=[\mathbf{x}\; \mathbf{y}\;
\mathbf{1}]$ with vectors of data points $\mathbf{x}, \mathbf{y}$, and
a vector of ones $\mathbf{1}$.  Note that the set of linear equations
(\ref{eq:vectorv}), even though derived by a different procedure, is
equivalent to the one obtained by K\'asa \cite{Kasa:1976} by a
modified least square criterion, \emph{i.e.,} by minimizing the sum
$\sum_i[(x_i-x_\mathrm{c})^2+(y_i-y_\mathrm{c})^2-R^2]^2$.  We would
like to remark that a great deal of research has been done on the
approximation of circular arcs and whole circles (see, \emph{e.g.,}
\cite{Jaklic:2007} and the references therein).  Since in our case the
vesicle contour is known to have circular shape, special polynomial or
spline approximations are not needed.

\begin{figure}
  \centering\includegraphics[scale=0.7]{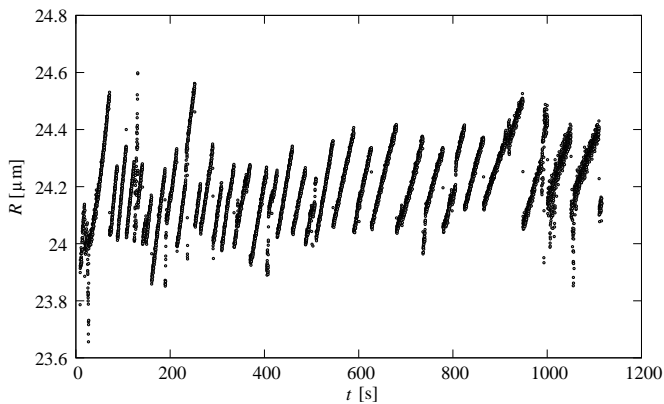}
  \caption{The determined vesicle radius $R$ as a function of time $t$
    upon a transfer from a 0.2~mol/L sucrose/glucose solution into an
    iso-osmolar glycerol solution.}
  \label{fig:sawtooth-data}
\end{figure}

A script was written in GNU Octave (\url{http://www.octave.org/},
\cite{Eaton:Octave}), a Matlab-compatible software package, to
determine the radius of the vesicle from a videomicrograph image.
Figure~\ref{fig:sawtooth-data} shows a series of calculated radii from
an 18-minute recording of a vesicle.  As verified by a visual
inspection, occasional apparent transient increases or decreases of
the radius are due to image defocussing (as the cover slide, along
which vesicles move, is not perfectly aligned with the focal plane,
the experimentalist needs to occasionally adjust the focussing knob on
the microscope in order to assure that the vesicle is still in
focus). Such transient changes in the apparent calculated vesicle
radius occur on an a time-scale of seconds and it is thus easy to
discriminate them agains real changes of vesicle radius.

Another GNU Octave script was used to read the output of the first
script, \emph{i.e.,} the vesicle radius as a function of time, and
determine the points at which vesicle bursts occurred.  The algorithm
employed was a simple one: the script looked for negative jumps in
vesicle radius, the magnitude of which exceeded some pre-determined
value and which did not follow the previous jump too soon.  Both
parameters---the minimum amplitude and the minimum time interval
between the jumps---were hand-tailored to fit the
experimental data.  Using this algorithm, 90--95\% of the jumps were
detected; the missing ones were added by hand upon comparing the
results to the data.

\section{Results and Discussion}
\label{sec:results}

\subsection{Test set of data}
\label{sec:test-set}

\begin{figure}
  \centering\includegraphics[scale=0.65]{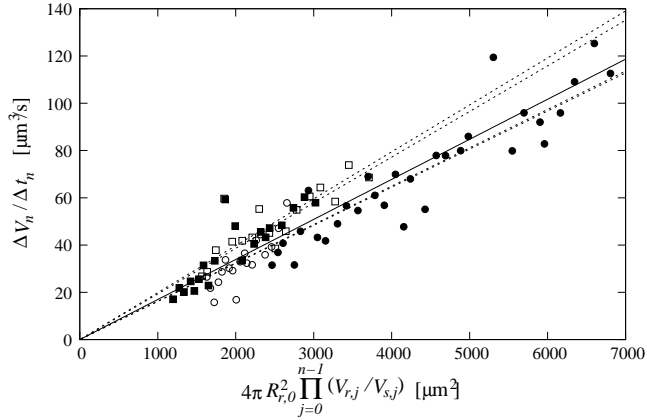}
  \caption{The rate of the volume increase $\Delta V_n / \Delta t_n$
    plotted against $4\pi R_\mathrm{r,0}^2 \prod_{j=0}^{n-1}
    V_{\mathrm{r},j}/V_{\mathrm{s},j}$, shown for four different
    vesicles transferred from 0.2~mol/L glucose/sucrose into 0.2~mol/L
    glycerol (denoted {\Large $\circ$} (19 points), {\Large $\bullet$}
    (31 points), $\square$ (17 points), and $\blacksquare$, 20
    points). Least-squares fitting of a straight line through the
    origin yields values of $P$ $(1.61\pm 0.07)\times 10^{-8}$~m/s,
    $(1.62\pm 0.04)\times 10^{-8}$~m/s, $(1.98\pm 0.07)\times
    10^{-8}$~m/s, and $(1.93\pm 0.08)\times 10^{-8}$~m/s, respectively
    (plotted with a dotted line; the lower two lines are barely
    distinguishable).  Here, the slope of a straight line through $n$
    points $(x_i,y_i)$ is calculated as $m=\sum_i x_iy_i/\sum_ix_i^2$,
    and its standard error as
    $\sum_i(y_i-mx_i)^2/((n-1)\sum_ix_i^2)$. A solid line fitted
    through all four sets of points corresponds to $P =
    (1.69\pm0.03)\times 10^{-8}\,\textrm{m/s}$.}
  \label{fig:permeability}
\end{figure}

The algorithm was tested on a set of data used in \cite{Peterlin:2008}
(figure~\ref{fig:permeability}).  Membrane permeability has been
calculated for four recorded series of bursts (vesicles 2--5 in
table~1, \cite{Peterlin:2008}), yielding an average value $P =
1.7\times 10^{-8}\,\textrm{m/s}$ at a room temperature
($26\pm2^\circ$C), with the estimates on individual vesicles ranging
from $1.61\times 10^{-8}\,\textrm{m/s}$ to $1.98\times
10^{-8}\,\textrm{m/s}$.  This value is lower than the value
$(2.09\pm0.82)\times 10^{-8}\,\textrm{m/s}$ obtained in
\cite{Peterlin:2008} by another method on the same data set, as well
as from other published data, \emph{e.g.,} the value for DOPC,
$2.75\times 10^{-8}\,\textrm{m/s}$ at $30^\circ$C \cite{Paula:1996},
and two values for egg-PC, $(4.3 \pm 0.1)\times 10^{-8}\,\textrm{m/s}$
\cite{Dordas:2000} and $5.4\times 10^{-8}\,\textrm{m/s}$ at
$25^\circ$C \cite{Orbach:1980}.

One has to note, however, that the only result against which we can
directly compare the obtained result is the one obtained by a
different analysis on the same data set \cite{Peterlin:2008}.  Both
results agree within the margins of standard error in
\cite{Peterlin:2008}.  All other results were obtained on systems
which were different, even though they share some similarities.  Paula
\emph{et al.} \cite{Paula:1996} did not include POPC in their
measurement.  The most similar system of those included is DOPC, which
has two unsaturated acyl chains (POPC has one unsaturated and one
saturated chain).  It is known that increasing chain saturation
decreases membrane fluidity, and membrane fluidity has been shown to
be well-correlated with the permeability of membrane for small
uncharged solutes \cite{Lande:1995}.  Furthermore, the experiment has
been conducted at a slightly higher temperature, and it is known that
membrane permeability scales with temperature approximately as
Arrhenius factor, $\propto \exp(-E_a/kT)$ \cite{Kleinhans:1998}.

The other two measurements \cite{Dordas:2000,Orbach:1980} were
conducted on natural egg phosphatidylcholine.  According to the
specifications provided by the producer (Avanti Polar Lipids, Inc.),
this is a mixture of various lipids, both saturated and unsaturated,
ranging in length of the acyl chain from 16 to 20 carbon atoms.  It is
to be expected that a heterogenous membrane has a higher permeability,
and indeed, the permeabilities of egg-PC membrane listed in
\cite{Dordas:2000,Orbach:1980} are consistently higher than those for
DOPC membrane listed in \cite{Paula:1996}.

As one can see, the presented method offers a possibility to determine
membrane permeability with a  relative accuracy
comparable to or greater than other published results.
Based on the above argumentation, one could expect that the value
obtained for permeability of POPC membrane should be slightly lower
than the published values for other similar systems, and indeed it is.
We have not shown, however, that the extent of the observed
discrepancy can be explained solely by the fact that we were working
with a different system.  We can think of several possible additional
factors which could also affect the estimated permeability.

Firstly, this discrepancy may indicate that the membrane permeability
depends on the membrane curvature and thus one may indeed expect
different values for the membrane of GUVs, which are essentially
planar, than for the membranes of LUVs and other smaller aggregates.
A comparison by Brunner \emph{et al.}  \cite{Brunner:1980} however
shows that planar membranes (black lipid membranes, BLM) exhibited
higher permeabilities for ions and sugars than the membranes of
sonicated small unilamellar vesicles (SUVs); it is to be noted,
though, that the authors report problems with leakage and attribute a
comparatively high value for BLM permeability to this artefact.

Secondly, it may be due to the fact that the model presented in
section \ref{sec:model} does not take into account the membrane
elasticity.  It has already been pointed out \cite{Chiruvolu:1993}
that failing to account for membrane elasticity in the interpretation
of the membrane permeability measurements in a similar experiment
leads to an underestimate of membrane permeability.  Our earlier
estimates \cite{Peterlin:2008} however show that this correction
amounts to only 2--3\% and thus cannot explain the observed
difference.

Thirdly, the apparent value of permeability, as obtained in the
experiment, is coupled with the diffusion of the permeable solute
through the stagnant layer \cite{Barry:1984} and is therefore lower
than the true permeability of the membrane alone.  A quick estimate
however shows that the effect of the stagnant layer only affects the
calculated permeability to a few percent.

Finally, one has to bear in mind that the phase contrast technique
does not yield the true position of the vesicle membrane.  Instead, we
have to rely on the ``halo'', which is an artefact of the technique,
and define the position of the membrane as the point of the maximal
gradient of the image intensity.  For a straight-edge object, it has
been calculated \cite{Wilson:1981} that the maximal steepness of the
profile slope indeed coincides with the position of the edge.  A
rigorous theoretical treatment would require repeating this
calculation for each vesicle, taking into account the real parameters
of the optical system used, which would considerably complicate the
analysis without a promise of a significant improvement.  As a
consequence, even the most precise analyses of vesicle contours to
date \cite{Dobereiner:1997,Pecreaux:2004} dismiss this discrepancy as
negligible and rely to the result of Wilson and Sheppard
\cite{Wilson:1981}.  All in all, we can conclude that the method
presented in this paper needs to be applied to a wider variety of
systems before this feature can be fully explained, and then possibly
applied to other interesting phenomena, such as the critical phenomena
predicted \cite{Haleva:2008}.

\subsection{Comparison with other halo-based techniques}

Although it is based on previous works
\cite{Mally:2002,Peterlin:2008}, the presented technique uses a
different method for determining the composition of the vesicle
interior, \emph{i.e.,} the amount of the impermeable solute
$N_\mathrm{i}^{(n)}$.  The previously published methods relied on the
height of the ``halo'' of the phase-contrast image for determining the
composition of the vesicle interior and tacitly assumed that all the
jumps of the vesicle radius were equal.
Figure~\ref{fig:sawtooth-data} demonstrates that this is not
necessarily the case.  Furthermore, since no \emph{ab initio}\/
calculation of the ``halo'' profile has been done, the authors in
\cite{Mally:2002} and \cite{Peterlin:2008} had to resort to the
calibration of the halo.  As the halo profile depends not only on the
refractive indexes of the solutions in the vesicle interior and the
vesicle exterior, but also on the size of the vesicle and, in absolute
terms, on the properties of the optical system used (\emph{e.g.,} a
slight displacement of the phase ring from the optical axis affects it
considerably), this analysis is cumbersome, and any laboratory trying
to apply this technique needs to repeat the calibration on their
system.  By replacing the need for the measurements of the halo
properties such as its height or width with a more precise
measurements of the vesicle radius, which enables us to estimate the
composition of the vesicle interior using simple geometry
(\ref{eq:linear}), we eliminated the need for the cumbersome
calibration altogether, and thus made the technique much more
portable.  A further advantage is that shedding the reliance on the
quantitative properties of the halo also opens the possibility for
using techniques for improving uneven illumination (see, \emph{e.g.,}
\cite{Likar:2000}), if necessary.  Note, however, that when using the
presented technique, it is important to record the vesicle behaviour
right from the time of the transfer, as missing a burst yields to
underestimating the permeability.

Applying the technique from \cite{Mally:2002} to the measurement of
small polar molecules, as has been done in \cite{Peterlin:2008},
introduces a further problem, since the calibration itself involves a
tricky part---the experimentalist has to determine a suitable point
for the initial concentration difference, \emph{i.e.,} a point late
enough that the amount of the initial solution transferred with the
micropipette along with the vesicle has diffused away enough not to
affect the refracting index of the external solution significantly,
and early enough that the amount of the permeable solute which has
entered the vesicle can still be neglected.  This procedure requires
some experience and can not be automated.  This does not present a
problem with the experiment where membrane permeability due to peptide
incorporation in the membrane is studied \cite{Mally:2002}, as the
permeability of membrane in the absence of peptides is very low and
the experimentalist has plenty of time for the calibration.  As the
technique presented here does not rely on the properties of the halo
profile, it provides a better alternative for measuring the
permeability of membrane for small polar molecules.

\section{Conclusions}

We describe a technique for determining the permeability of a
phospholipid membrane of micrometre-sized GUVs from a sequence of
videomicrographs, which complements the existing methods of
determining membrane permeability on either sub-micrometre-sized
liposomes (SUV, LUV) or planar membranes.  The technique relies on
determining the radius of a spherical GUV with a great precision, and
offers a substantial improvement over similar techniques which rely on
quantifying the properties of the phase-contrast halo such as the
height or width of its profile.  When applied to a test set of data,
the technique yielded consistent results.  The obtained value is,
however, $\approx 40\%$ lower than the results obtained on similar
systems.  Further investigations on a system comparable to the ones in
the literature are needed to determine whether this discrepancy would
still persist.

\ack{The authors would like to thank V.~Arrigler for her skillful help
  with phospholipid vesicles, and M.~Rai\v{c} and M.~Pisanski
    for a helpful discussion.  This work has been supported by the
  Slovenian Research Agency through grants P1-0055 (PP) and P1-0294
  (GJ, TP).}

\end{document}